\documentclass{article}
\usepackage{ijcai26}

\usepackage{times}
\usepackage{soul}
\usepackage{url}
\usepackage[hidelinks]{hyperref}
\usepackage[utf8]{inputenc}
\usepackage[small]{caption}
\usepackage{graphicx}
\usepackage{amsmath}
\usepackage{amsthm}
\usepackage{booktabs}
\usepackage{algorithm}
\usepackage{algorithmic}
\usepackage[switch]{lineno}

\usepackage{booktabs}
\usepackage{multirow}

\usepackage[framemethod=tikz]{mdframed}
\usepackage{enumitem}
\usepackage{listings}
\usepackage{graphicx}
\usepackage{amsmath}
\usepackage{amssymb}
\usepackage{subcaption}
\usepackage{verbatim}

\title{Do We Really Need SFT? Prompt-as-Policy over Knowledge Graphs for Cold-start Next POI Recommendation}

\author{
    Anonymous Author 
     \affiliations
     PaperID: 1267
}

\author{
Jinze Wang$^{1,2,*}$
\and
Lu Zhang$^{3,*}$\and
Tiehua Zhang$^{1,\dag}$\and
Yiyang Cui$^{1}$\and
Zhishu Shen$^4$\and
Yuze Liu$^2$\and
Xingjun Ma$^5$\and
Jiong Jin$^2$
\\
\affiliations
$^1$Tongji University\\
$^2$Swinburne University of Technology\\
$^3$Chengdu University of Information Technology\\
$^4$Wuhan University of Technology\\
$^5$Fudan University\\
\emails
tiehuaz@tongji.edu.cn
}

\begin{document}

\maketitle

\begin{abstract}

Next point-of-interest (POI) recommendation is a key component of smart urban services, yet it remains challenging under cold-start conditions with sparse user–POI interactions.
Recent LLM-based methods address this issue through either supervised fine-tuning (SFT) or in-context learning (ICL), but SFT is costly and prone to overfitting active users, while static prompts in ICL lack adaptability to diverse user contexts.
We argue that the main limitation lies not in LLM reasoning ability, but in how contextual evidence is constructed and presented.
Accordingly, we propose Prompt-as-Policy over knowledge graphs (KG), a reinforcement-guided prompting framework that formulates prompt construction as a learnable decision process, while keeping the LLM frozen as a reasoning engine.
To enable structured prompt optimization, we organize heterogeneous user–POI signals into a KG and transform mined relational paths into evidence cards, which serve as atomic semantic units for prompt composition.
A contextual bandit learner then optimizes a prompt policy that adaptively determines
(i) which relational evidences to include,
(ii) how many evidences to retain per candidate POI, and
(iii) how to organize and order them within the prompt.
Experiments on three real-world datasets show that Prompt-as-Policy consistently outperforms state-of-the-art baselines, achieving an average 11.87\% relative improvement in Acc@1 for inactive users, while maintaining competitive performance for active users, without any model fine-tuning.

\end{abstract}

\section{Introduction}

Location-based social networks (LBSNs) and mobile applications have made next point-of-interest (POI) recommendation an indispensable component of smart urban services, supporting applications from tourism and dining to transportation and retail~\cite{cui2021sequential,sun2021point,11210020,wu2025decentralized}. Traditional approaches, particularly graph-based methods have significantly advanced the ability to capture spatial–temporal dependencies and user mobility patterns~\cite{ijcai2021p206,velivckovic2018graph}. Despite these successes, they still struggle under cold-start conditions, where limited interaction makes it challenging to infer reliable user preferences and movement intentions. As a result, achieving accurate recommendations for inactive users remains an open problem.

Recently, large language models (LLMs) have shown exceptional potential as reasoning engines in recommendation systems, owing to their ability to integrate commonsense knowledge with contextual information through natural language prompts~\cite{10.1145/3604915.3610647,ma2025pub}. 
Most existing LLM-based recommendation frameworks rely on supervised fine-tuning (SFT), which requires extensive annotated data and considerable computational resources. 
Moreover, SFT models tend to overfit users with rich interaction histories, leading to weak generalization in cold-start scenarios~\cite{lyu-etal-2024-llm}. 
To alleviate these issues, a growing body of research has explored in-context learning (ICL) approaches, which encode task instructions directly into static prompts~\cite{li2024large,wongso2024genup}. 
While these methods avoid costly fine-tuning, they are inherently limited by their fixed prompt templates, which cannot dynamically adapt to variations in user contexts. Consequently, both SFT and static-prompt paradigms exhibit limited scalability and robustness in real-world cold-start environments.

Additionally, some LLM-based studies focus on prompt optimization and reveal that the reasoning behavior is highly sensitive to variations in prompt composition and ordering, underscoring that how prompts are constructed can be as crucial as what information they contain~\cite{liu2021makes,lu2022fantastically,liu2025grl,ashizawaetal2025bandit}. This phenomenon suggests that improving the prompting process itself, rather than relying solely on SFT, may unlock generalizable reasoning capabilities.

\textcolor{black}{
Motivated by these observations, we revisit the role of prompting in LLM-based recommendation and argue that prompt construction itself should be treated as a first-class optimization problem, rather than a fixed design choice.
We thus propose Prompt-as-Policy, a reinforcement-guided prompting framework that formulates prompt construction as a learnable decision-making process.
Instead of fine-tuning model parameters or relying on static prompt templates, we keep the LLM frozen as a reasoning engine and focus on optimizing how contextual information is presented to the model. }

\textcolor{black}{
To make prompt optimization tractable and controllable, we first structure the raw user–POI interaction signals into an explicit evidence space. Specifically, we construct a knowledge graph (KG) that encodes heterogeneous relations among users, POIs, categories, spatial grid cells, time slots, intents, and profile anchors. From this KG, we extract user–POI relational paths and transform them into evidence cards, each of which summarizes a concise and interpretable relational rationale (e.g., user intent alignment or spatial proximity). These evidence cards serve as the atomic semantic units from which prompts are composed.}

\textcolor{black}{
Under this formulation, prompt construction is no longer a monolithic template but a structured configuration problem over a set of candidate evidences. A contextual bandit learner is therefore introduced to learn a prompt policy that adaptively selects and organizes evidence cards based on the user context and candidate POIs. 
Concretely, the policy optimizes three key dimensions of the prompting process:
(i) which relational evidences to include,
(ii) how many rationales to retain for each candidate POI, and
(iii) how to organize and order these rationales within the prompt.
By optimizing these dimensions jointly through reinforcement signals, Prompt-as-Policy enables adaptive, context-aware prompting that improves robustness and generalization in cold-start recommendation scenarios.}


\textbf{Contributions.} To the best of our knowledge, we are the first to formulate prompt construction as a learnable policy for LLM–based recommendation for cold-start scenario. The main contributions are summarized as follows:
\begin{enumerate}
  \item We propose Prompt-as-Policy framework, a reinforcement guided prompting framework that replaces SFT with dynamic prompt optimization. It integrates knowledge-graph based relational path mining, evidence cards, and RL-based adaptive prompt learner under cold-start conditions.
  \item We formulate prompt construction as a contextual bandit optimization problem, where the learned policy adaptively determines \textit{which relational evidences to include, how many rationales to retain, and how to organize and order them within prompts.} 
  \item We conduct extensive experiments on three real-world Foursquare city datasets, covering different user activity levels and evidence-card configurations. Ablation studies further validate each component’s effectiveness, confirming the superiority of Prompt-as-Policy under various cold-start conditions.
\end{enumerate}
\section{Related Work}
\subsection{Next POI Recommendation}
Next-POI recommendation has garnered considerable attention in recent years, fueled largely by rapid progress in deep learning methodologies. Most existing approaches rely on sequential modeling, where user trajectories are treated as ordered sequences to capture temporal dependencies and behavioral dynamics~\cite{liu2016predicting,zhao2020go}. However, such methods often overlook latent spatial connectivity, which is crucial for representing complex mobility patterns~\cite{zhang2022next,zhang2025kg4receval}. The emergence of graph neural networks (GNNs) has opened new avenues for next-POI recommendation by enabling explicit modeling of spatial and temporal correlations through structured relational graphs~\cite{velivckovic2018graph,10.1145/3397271.3401063,ijcai2021p206,lei2025context}. Nevertheless, most GNN-based methods primarily model pairwise relations and may fail to capture higher-order dependencies among multiple POIs and contextual factors. Although knowledge graph and hypergraph learning have been introduced to alleviate this limitation by incorporating richer relational structures~\cite{10.1016/j.knosys.2022.109951,yan2023spatio,luo2025hypergraphrag}, these methods still struggle under cold-start conditions, where limited user--POI interactions make it difficult to infer reliable user preferences and mobility intentions. Consequently, generating accurate recommendations for users with sparse histories remains an open and challenging problem.

\subsection{LLMs for Next-POI Recommendation}

LLMs have emerged as powerful reasoning engines for user mobility prediction~\cite{10.1145/3604915.3610647,lyu-etal-2024-llm}, and existing studies generally follow two main directions: supervised fine-tuning and in-context learning.

\noindent\textbf{Supervised Fine-Tuning (SFT).} Recent studies employ SFT to explicitly align LLM parameters with user--POI interactions, thereby improving domain relevance and predictive accuracy for next-POI recommendation. For instance, LLM4POI~\cite{li2024large} captures user mobility patterns by fine-tuning LLMs with trajectory similarity signals. GenUP~\cite{wongso2024genup} extends this line of work by replacing long historical trajectories with generative natural-language user profiles that summarize preferences, routines, and personality traits inferred from check-ins. GNPR-SID~\cite{10.1145/3711896.3736981} adopts SFT with semantically constructed POI IDs for a generative next-POI recommendation framework. GA-LLM~\cite{liu2025geography} further integrates geographic context and sequential transition modeling into fine-tuned LLMs for next-POI recommendation. Despite these advances, SFT-based approaches typically require substantial training data and computational resources, or rely on rich side information (e.g., geographic context), which limits their practicality in cold-start settings.

\noindent\textbf{In-Context Learning (ICL).} 
Recent studies have turned to ICL, where prompts are constructed to guide models toward more accurate outputs.
LLM-Mob~\cite{wang2023would} forms and engineers context-inclusive prompts to enable LLMs to reason over mobility contexts and generate accurate predictions. PromptRec~\cite{wu2024could} provides a mathematical framework to study how language models make in-context recommendations without any parameter updates. RALLM-POI~\cite{li2025rallm} integrates retrieval-based augmentation and geographical re-ranking to enhance zero-shot performance. LightPROF~\cite{ao2025lightprof} incorporates structured KG information into prompts to guide reasoning through multi-hop relational evidence. Refine-POI~\cite{li2025refine} adopts a reinforcement-based framework and develops location-aware trajectory prompting to better leverage geographic knowledge.


\textcolor{black}{Nevertheless, existing methods either rely on static prompts or require costly retraining, and fail to adaptively exploit relational knowledge in user mobility data.
To overcome these limitations, we propose Prompt-as-Policy, a reinforcement-guided prompting framework that learns a prompt construction policy over mined relational evidences via contextual bandit optimization.
By dynamically selecting and organizing relational evidences, our method enables a frozen LLM to generate adaptive recommendations under cold-start conditions without any fine-tuning.}

\begin{figure*}[ht]  
    \includegraphics[width=1\textwidth]{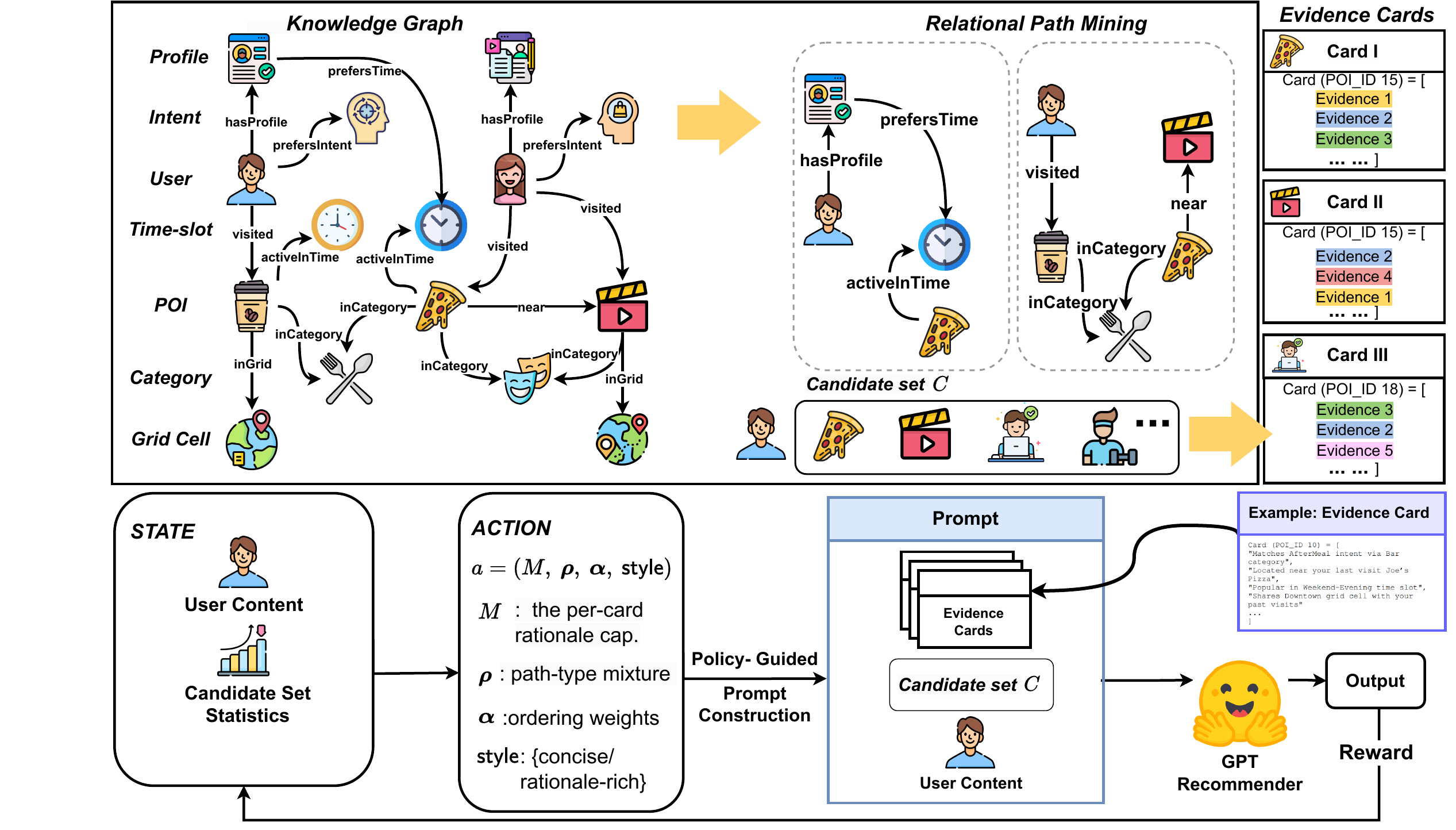}
    \caption{Overview of the proposed \textbf{Prompt-as-Policy} framework.
    }
     \label{fig:framework}
\end{figure*}

\section{Prompt-as-Policy}
\subsection{Problem Formulation}
Let $\mathcal{U}$ and $\mathcal{P}$ denote the sets of users and POIs. For a user $u\in\mathcal{U}$, the historical check-ins are $\mathcal{S}_u=\{(p_i,\gamma_i)\}_{i=1}^{|\mathcal{S}_u|}$ in chronological order, where $p_i\in\mathcal{P}$ and timestamp $\gamma_i\in\mathbb{R}$. At recommendation time we construct the user context $\mathbf{x}=(\delta,\ell,F_u,I_u)$, where $\delta$ is the user visit time-slot (e.g., morning/afternoon/evening/night), $\ell$ is the most recent location (last check-in), $F_u$ summarizes user profile statistics (e.g., top categories, hotspot grids, mobility radius), and $I_u$ is the current behavioral intent inferred by an LLM reasoning engine (e.g., afterMeal, social, relax, shopping). Under cold-start conditions (e.g., inactive users), the task is to predict the next POI $p^*$ from a candidate set $\mathcal{C}$. It is worth noting that we \textbf{do not} fine-tune the LLM (no SFT) and \textbf{do not} train KG embeddings. Instead, all learning targets a prompt policy that decides how to select and organize evidence for the LLM.

\subsection{Overall Framework}
The overall Prompt-as-Policy framework of our work is presented in Figure~\ref{fig:framework}. First, we construct the KG to discover candidate POIs and extract relational paths that represent user–POI correlations. Then, we perform path mining and candidate discovery to explore relational paths from users to POIs and form the candidate set. Next, we conduct evidence mining to summarize these paths into interpretable evidence cards. After that, prompt construction integrates the user context, candidate list, and evidence cards into a structured input. Finally, a policy learner based on contextual bandit reinforcement learning optimizes the prompt policy, adaptively selecting and organizing evidence before the constructed prompt is fed into the LLM for next POI recommendation.

\subsection{Knowledge Graph}

We construct a heterogeneous knowledge graph $\mathcal{G} = (\mathcal{V}, \mathcal{E}, \mathcal{R})$ to (i) discover candidate POIs via relational path mining and (ii) extract structured evidence for prompt construction. The entity set $\mathcal{V}$ contains seven types, while edges represent typed relations $\mathcal{R}$ connecting these entities. Table~\ref{tab:kg_schema} summarizes the schema for the construction of KG. Specifically, each edge is a directed triple $(h, r, v) \in \mathcal{V} \times \mathcal{R} \times \mathcal{V}$, where $r$ denotes the type of relation.

\begin{table}[h]
\small
\centering
\begin{tabular}{ll|ll}
\toprule
\multicolumn{2}{c|}{\textbf{Entity Types}} & \multicolumn{2}{c}{\textbf{Relation Types}} \\
\midrule
$\mathcal{V}_U$ & Users & \textit{visited} & User → POI \\
$\mathcal{V}_P$ & POIs & \textit{hasProfile} & User → Profile \\
$\mathcal{V}_C$ & Categories & \textit{prefersIntent} & Profile → Intent \\
$\mathcal{V}_G$ & Grid cells & \textit{prefersTime} & Profile → TimeSlot \\
$\mathcal{V}_T$ & Time slots & \textit{inCategory} & POI → Category \\
$\mathcal{V}_I$ & Intents & \textit{inGrid} & POI → Grid \\
$\mathcal{V}_F$ & Profile anchors & \textit{activeInTime} & POI → TimeSlot \\
 & & \textit{near} & POI → POI \\
\bottomrule
\end{tabular}
\caption{Knowledge graph construction schema.}
\label{tab:kg_schema}
\end{table}

Grid cells $\mathcal{V}_G$ are obtained by spatial clustering with $k$-means on POI coordinates. The \textit{near} relation connects POIs within a Haversine distance threshold $d_{\text{near}}$. Following empirical findings that urban mobility transitions typically occur within 10km~\cite{wang2023meta}, we set 
$d_{\text{near}} = 10$km across all datasets to capture realistic movement patterns while maintaining computational efficiency. The \textit{activeInTime} relation links each POI to time slots during which it received historical check-ins. For path traversal, we allow bidirectional edge following to ensure connectivity, while keeping the rationale verbalization consistent with the canonical semantics. When traversing reversed edges, we only use them to maintain graph connectivity, while the evidence rationales always verbalize relations in their canonical semantic direction.

\subsection{Relational Path Mining}
Instead of selecting candidates only by proximity, we employ a graph traversal strategy to mining candidates from the KG by executing a multi-hop breadth-first search (BFS) originating from the user node $u$ within a limited number of hops. This constrained traversal naturally explores the KG to discover POIs via diverse relational paths, effectively capturing semantic patterns such as intent correlation ($U\rightarrow F\rightarrow I\rightarrow C\rightarrow P$), grid collaborative signals ($U\rightarrow P \rightarrow G \rightarrow P$), spatial proximity ($U\rightarrow P\rightarrow \textit{near}\rightarrow P$), and temporal preferences ($U\rightarrow F \rightarrow T\rightarrow P $). By prioritizing search breadth, BFS ensures that shorter, more interpretable paths are discovered before longer ones, which is crucial for generating concise evidence rationales. To ensure candidate quality, we apply distance checks to discard POIs whose distance from the last check-in $\ell$ exceeds a city-scale threshold $\bar D$. After filtering, the remaining POIs are deduplicated to form the candidate set $\mathcal{C}$; for interaction round $t$, we denote the resulting candidate set instance as $C_t$.

\noindent\textbf{Evidence Cards.} Given the final candidate set $\mathcal{C}$, we mine a set of valid paths by enumerating instance paths from $u$ to each $p'\in\mathcal{C}$ discovered during the traversal. Each remaining path $q$ is summarized into a one-line rationale $\rho=\textsf{summarize}(q)$ by an LLM following a predefined template. For each candidate $p'$, we collect a pool of rationales by summarizing all mined evidence paths, $\textsf{Pool}(p')=\{\rho_1,\ldots,\rho_{n(p')}\}$, where $n(p')$ is the number of summarized rationales for $p'$. The prompt policy then selects and orders up to $M$ rationales to form the final evidence card, $\textsf{Card}(p')=[\,p' : \{\rho'_1,\ldots,\rho'_{m(p')}\}\,]$, where $m(p') = \min\{M, n(p')\}$ and the rationales $\{\rho'_k\}$ are selected from $\textsf{Pool}(p')$ by the policy. Examples of evidence cards refer to Appendix A.2.

The policy learner introduced in Section~\ref{sec:promptconstruction} will decide which rationales to keep from the pool, how many to use via $M$, and how they are ordered. The prompt policy not only determines the subset and quantity of evidences but also controls their presentation sequence. Recent research has shown that the reasoning performance of LLMs can be highly sensitive to variations in in-context examples~\cite{liu2025grl,lu2022fantastically,liu2021makes}, motivating our design to explicitly model evidence ordering as part of the policy. 

\subsection{Policy-Guided Prompt Construction.}\label{sec:promptconstruction} The prompt presented to the LLM has three components: (i) a compact user context header ($\delta,\ell$, a brief profile summary $F_u$, and the intents $I_u$); (ii) the KG-discovered candidate list containing $(\text{id},\text{category},\text{distance})$ for all $p\in C_t$; and (iii) the evidence cards, one per candidate, each with up to $M$ policy-selected rationales. It is worth noting that the user context is explicitly integrated into the system prompt to guide the LLM’s reasoning process. Following recent findings~\cite{wongso2024genup}, such integration improves the stability and consistency of LLM outputs by providing a persistent contextual anchor. The LLM is used as a reasoning engine and must output a strict JSON object \texttt{\{"ranking":[...]\}} whose IDs belong to $C_t$ only. Any schema violation or out-of-candidate ID triggers a penalty, which is incorporated into the reward design in the next section. Examples of prompt construction refer to Appendix A.4.

\noindent\textbf{Policy Learner.} Motivated by the fact that prompt construction is a single-step decision process with immediate feedback~\cite{feng2024move,ashizawa2025bandit}, we cast prompt construction as a contextual bandit problem and learn the policy with contextual-bandit reinforcement learning instantiated as Contextual Thompson Sampling (CTS). In our setting, each training interaction yields a single immediate reward after one prompt construction, so there is no delayed or long horizon credit assignment and a bandit formulation is sufficient.

We index bandit rounds by $t$, where each round corresponds to one next POI prediction decision on the training split. At round $t$, the environment reveals a state
\begin{equation}
s_t = \bigl(x_t, \phi(C_t)\bigr),
\end{equation}
where $x_t$ is the user context and $\phi(C_t)$ summarizes candidate statistics. The action space consists of prompt configurations
\begin{equation}
a = \bigl(M, \boldsymbol{\omega}, \boldsymbol{\alpha}, \text{style}\bigr),
\end{equation}
where $M$ is the per-card rationale cap ($1$--$N$), $\boldsymbol{\omega}$ governs the preference weights for different path types (e.g., prioritizing intent vs. spatial signals), $\boldsymbol{\alpha}$ sets rationale ordering weights (e.g., relevance-first, diversity-first), and $\text{style} \in \{\text{concise}, \text{rationale-rich}\}$ controls textual verbosity. In practice, each of these dimensions is discretized into a small set of options, and their Cartesian product forms a finite action set $\mathcal{A}$.

To apply CTS, we follow a standard contextual bandit parameterization. For each state and action pair $(s_t,a)$ we construct a $d$-dimensional feature vector
\begin{equation}
\mathbf{z}(s_t,a) \in \mathbb{R}^d,
\end{equation}
where the action components are encoded as discrete indicator vectors and concatenated with normalized numerical state statistics. We assume a linear Gaussian reward model
\begin{equation}
R_t = \boldsymbol{\theta}^{\top} \mathbf{z}(s_t,a_t) + \varepsilon_t,
\qquad
\varepsilon_t \sim \mathcal{N}(0,\sigma^2),
\end{equation}
with a Gaussian prior $\boldsymbol{\theta} \sim \mathcal{N}(\boldsymbol{\mu}_0, \mathbf{\Sigma}_0)$. Following the common implementation of linear Thompson Sampling, we maintain
\begin{equation}
\mathbf{A}_t \in \mathbb{R}^{d \times d},
\qquad
\mathbf{b}_t \in \mathbb{R}^d,
\end{equation}
initialized as $\mathbf{A}_0 = \lambda_{\text{reg}} \mathbf{I}_d$ and $\mathbf{b}_0 = \mathbf{0}$, where $\lambda_{\text{reg}} > 0$ is a regularization parameter and $\mathbf{I}_d$ is the $d$-dimensional identity matrix. 
At round $t$, we first compute the posterior mean
\begin{equation}
\widehat{\boldsymbol{\theta}}_t = \mathbf{A}_{t-1}^{-1} \mathbf{b}_{t-1},
\end{equation}
and sample a parameter
\begin{equation}
\widetilde{\boldsymbol{\theta}}_t \sim \mathcal{N}\bigl(\widehat{\boldsymbol{\theta}}_t, \nu^2 \mathbf{A}_{t-1}^{-1}\bigr),
\end{equation}
where $\nu > 0$ controls the exploration level. Given $s_t$, CTS then selects the action
\begin{equation}
a_t = \arg\max_{a \in \mathcal{A}} \widetilde{\boldsymbol{\theta}}_t^{\top} \mathbf{z}(s_t,a),
\end{equation}
we then select and order each evidence card according to $a_t$, construct the prompt, obtain the LLM ranking constrained to the candidate set $C_t$, compute the reward $R_t = R(s_t,a_t)$ defined in the next paragraph, and update
\begin{equation}
\mathbf{A}_t = \mathbf{A}_{t-1} + \mathbf{z}(s_t,a_t)\mathbf{z}(s_t,a_t)^{\top},
\end{equation}
\begin{equation}
\qquad
\mathbf{b}_t = \mathbf{b}_{t-1} + R_t \,\mathbf{z}(s_t,a_t).
\end{equation}

This procedure is repeated over all training interactions. On the validation and test splits, the learned policy (i.e.,  $\mathbf{A}_t$ and $\mathbf{b}_t$)
is frozen and no further updates. The policy thus adaptively determines which relational evidences to include via $\boldsymbol{\omega}$, how many rationales to retain per candidate via $M$ and $\text{style}$, and how to organize and order them via $\boldsymbol{\alpha}$, while the KGs and the LLM remain fixed throughout.

\noindent\textbf{Reward Design.} We optimize a scalar reward that directly reflects the effectiveness of a prompt configuration for next POI prediction under the action space. Given $(s_t,a_t)$, we construct the prompt as described in Section~\ref{sec:promptconstruction} and obtain from the LLM a ranked list over the candidate set $C_t$ for round $t$,
\begin{equation}
\hat{\mathbf{y}}_t = \bigl(\hat{p}_{t,1}, \hat{p}_{t,2}, \dots, \hat{p}_{t,|C_t|}\bigr),
\end{equation}
where each $\hat{p}_{t,k} \in C_t$. Let $p_t^{\ast}$ denote the ground-truth POI for this interaction. We define the Top-$K$ accuracy signal as
\begin{equation}
\mathrm{Acc@K}\bigl(\hat{\mathbf{y}}_t, p_t^{\ast}\bigr)
=
\mathbb{I}\bigl[p_t^{\ast} \in \{\hat{p}_{t,1},\dots,\hat{p}_{t,K}\}\bigr],
\end{equation}
where $\mathbb{I}[\cdot]$ is the indicator function. We further define a validity indicator for the LLM output $o_t$ as
\begin{equation}
\mathrm{valid}(o_t, C_t)
= \mathbb{I}\!\left[
\substack{
o_t \text{ follows the JSON schema} \\
\text{and all IDs in } o_t \in C_t
}
\right].
\end{equation}

Let $L_t$ denote the prompt length (number of tokens) under $(s_t,a_t)$. Schema validity and token budget are treated as hard constraints: if the generated output violates the JSON schema, uses POI IDs outside $C_t$, or exceeds a token budget $\tau$, we assign zero reward. Formally, the reward is defined as
\begin{equation}
R(s_t,a_t) =
\begin{cases}
0, & \text{if } \mathrm{valid}(o_t, C_t) = 0 \\
    & \text{or } L_t > \tau, \\[5pt]
\mathrm{Acc@K}(\hat{\mathbf{y}}_t, p_t^{\ast}), & \text{otherwise.}
\end{cases}
\end{equation}

We set $K = 5$ to align the reward with the $\mathrm{Acc@K}$ evaluation metric, ensuring that the reward directly reflects end-task performance under schema and token-budget constraints. 
The effects of different prompt configurations, i.e., evidence selection, rationale budget, and ordering, are implicitly captured through their impact on Acc@K.

\section{Experiments}

\begin{table*}[th]
\centering
\fontsize{9}{9.5}\selectfont
\setlength{\tabcolsep}{3.5pt}
\renewcommand{\arraystretch}{1.15}
\begin{tabular}{cc|ccccccccccc}
\toprule
City & Metric
& STHGCN
& LLMMob
& PromptRec
& LLM4POI
& LightPROF
& GenUP
& RefinePOI
& GNPR-SID
& GA-LLM
& \textbf{Ours} \\
\midrule

\multirow{4}{*}{NYC}
& Acc@1   & 0.2734 & 0.2740 & 0.2866 & 0.2575 & 0.3409 & 0.3372 & 0.3469 & 0.3618 & \underline{0.3919} & \textbf{0.4023} \\
& Acc@5   & 0.5124 & 0.5083 & 0.5217 & 0.4896 & 0.6152 & 0.6085 & 0.5984 & 0.6421 & \underline{0.6754} & \textbf{0.6892} \\
& NDCG@5  & 0.3956 & 0.3914 & 0.4052 & 0.3728 & 0.4825 & 0.4764 & 0.4812 & 0.5083 & \underline{0.5412} & \textbf{0.5537} \\
& MRR     & 0.3582 & 0.3557 & 0.3684 & 0.3351 & 0.4416 & 0.4358 & 0.4520 & 0.4632 & \underline{0.4925} & \textbf{0.5041} \\
\cmidrule(lr){1-2}

\multirow{4}{*}{TKY}
& Acc@1   & 0.2592 & 0.2653 & 0.2735 & 0.2442 & 0.3174 & 0.3035 & 0.3160 & 0.3062 & \underline{0.3482} & \textbf{0.3565} \\
& Acc@5   & 0.4853 & 0.4927 & 0.5054 & 0.4658 & 0.5856 & 0.5623 & 0.5140 & 0.5714 & \underline{0.6258} & \textbf{0.6384} \\
& NDCG@5  & 0.3754 & 0.3812 & 0.3926 & 0.3554 & 0.4557 & 0.4362 & 0.4185 & 0.4418 & \underline{0.4935} & \textbf{0.5048} \\
& MRR     & 0.3387 & 0.3456 & 0.3552 & 0.3205 & 0.4153 & 0.3945 & 0.4020 & 0.3987 & \underline{0.4462} & \textbf{0.4573} \\
\cmidrule(lr){1-2}

\multirow{4}{*}{CAL}
& Acc@1   & 0.2341 & 0.2465 & 0.2523 & 0.2298 & 0.2876 & 0.2943 & 0.3052 & 0.3125 & \underline{0.3289} & \textbf{0.3412} \\
& Acc@5   & 0.4482 & 0.4654 & 0.4786 & 0.4357 & 0.5352 & 0.5486 & 0.5613 & 0.5723 & \underline{0.5945} & \textbf{0.6124} \\
& NDCG@5  & 0.3426 & 0.3583 & 0.3685 & 0.3324 & 0.4158 & 0.4253 & 0.4395 & 0.4467 & \underline{0.4658} & \textbf{0.4815} \\
& MRR     & 0.3084 & 0.3256 & 0.3327 & 0.3015 & 0.3782 & 0.3854 & 0.3967 & 0.4056 & \underline{0.4231} & \textbf{0.4376} \\
\midrule

\multicolumn{2}{l|}{No-SFT}
 & - & $\checkmark$ & $\checkmark$ & $\times$ & $\checkmark$ & $\times$ & $\checkmark$ & $\times$ & $\times$ & $\checkmark$ \\
\multicolumn{2}{l|}{Policy-guided Prompt}
 & - & $\times$ & $\times$ & $\times$ & $\times$ & $\times$ & $\times$ & $\times$ & $\times$ & $\checkmark$ \\
\bottomrule
\end{tabular}
\caption{Performance comparison on three datasets, where $\checkmark$ and $\times$ indicate the use of SFT and policy-guided prompt construction, respectively, and $-$ denotes non-LLM baselines. The best results are highlighted in \textbf{bold}, and the runner-up results are \underline{underlined}.}
\label{tab:overall_performance}
\end{table*}

\begin{table*}[th]
\centering
\fontsize{9.5}{9.5}\selectfont
\renewcommand{\arraystretch}{1.15}
\setlength{\tabcolsep}{3.5pt}
\begin{tabular}{lccccccccc}
\toprule
\multirow{3}{*}{Model} & \multicolumn{3}{c}{NYC} & \multicolumn{3}{c}{TKY} & \multicolumn{3}{c}{\textbf{CAL}} \\
\cmidrule(lr){2-4} \cmidrule(lr){5-7} \cmidrule(lr){8-10}
 & \textbf{Inactive} & \textbf{Normal} & \textbf{Active} & \textbf{Inactive} & \textbf{Normal} & \textbf{Active} & \textbf{Inactive} & \textbf{Normal} & \textbf{Active} \\
 & \scriptsize{$\bar{L}=1.9$} & \scriptsize{$\bar{L}=6.9$} & \scriptsize{$\bar{L}=26.5$} & \scriptsize{$\bar{L}=2.8$} & \scriptsize{$\bar{L}=10.1$} & \scriptsize{$\bar{L}=34.0$} & \scriptsize{$\bar{L}=2.4$} & \scriptsize{$\bar{L}=13.8$} & \scriptsize{$\bar{L}=30.8$} \\
\midrule
LightPROF          & 0.2014 & 0.3150 & 0.3715 & 0.1685 & 0.2915 & 0.3352 & 0.1520 & 0.2825 & 0.3225 \\
GenUP           & 0.1625 & 0.3015 & 0.3652 & 0.1452 & 0.2854 & 0.3285 & 0.1385 & 0.2715 & 0.3154 \\
Refine-POI         & \underline{0.2245} & 0.3312 & 0.3824 & \underline{0.1952} & 0.2985 & 0.3456 & \underline{0.1885} & 0.2895 & 0.3328 \\
GNPR-SID           & 0.1855 & 0.3425 & 0.4056 & 0.1652 & 0.3056 & 0.3585 & 0.1585 & 0.2952 & 0.3482 \\
GA-LLM             & 0.1982 & \underline{0.3856} & \underline{0.4325} & 0.1756 & \underline{0.3345} & \underline{0.3852} & 0.1695 & \underline{0.3156} & \underline{0.3654} \\
\midrule
\textbf{Ours}      & \textbf{0.2515} & \textbf{0.3925} & \textbf{0.4358} & \textbf{0.2185} & \textbf{0.3412} & \textbf{0.3895} & \textbf{0.2105} & \textbf{0.3225} & \textbf{0.3695} \\
\textit{Improv.}   & \textit{+12.02\%} & \textit{+1.79\%} & \textit{+0.76\%} & \textit{+11.93\%} & \textit{+2.00\%} & \textit{+1.12\%} & \textit{+11.67\%} & \textit{+2.18\%} & \textit{+1.12\%} \\
\bottomrule
\end{tabular}
\caption{User cold-start analysis on inactive, normal and very active users across three datasets. The average trajectory length ($\bar{L}$) for each group is indicated in the header. The best results are highlighted in \textbf{bold}, and the runner-up results are \underline{underlined}.}
\label{tab:coldstart_performance}
\label{tab:user_group_detailed}
\end{table*}


Our experiments evaluate the proposed Prompt-as-Policy framework by addressing the following questions: 
\textbf{(RQ1)} Can Prompt-as-Policy outperform existing baselines for next POI recommendation under identical experimental settings?
\textbf{(RQ2)} How does Prompt-as-Policy perform in user cold-start scenarios compared with existing methods?
\textbf{(RQ3)} How does the learned prompt policy improve recommendation performance through evidence selection, ordering, and cardinality control?
\textbf{(RQ4)} How do different components of Prompt-as-Policy contribute to the overall performance gains, and how consistent are the results across different LLM backbones?

\subsection{Experimental Setup}


\noindent\textbf{Datasets and Implementation Details.} We evaluate our approach on three widely used Foursquare datasets~\cite{yang2013sentiment}, namely NYC, CAL, and TKY, covering approximately 11 months from April 2012 to February 2013 and consisting of user check-ins from the Foursquare platform. 
Following the preprocessing procedure described in~\cite{li2024large,wongso2024genup}, each user’s check-in sequence is segmented into trajectories using a 24-hour sliding window, and trajectories containing only a single check-in are removed. The check-ins are ordered chronologically and divided into training, validation, and test sets, with the first 80\% used for training, the next 10\% for validation, and the remaining 10\% for testing. We use the same lightweight inference-only LLM, \texttt{gpt-4o-mini}, for the reasoning, user-intent inference, and rationale summarization. Additional implementation details and hyperparameter settings are provided in Appendix A.3. The source code is available at this \href{https://github.com/Prompt-as-Policy/IJCAI}{\textbf{Code link}}.


\noindent\textbf{Baseline.} We compare our method with three categories of SOTA baselines:
(i) Graph-based models such as STHGCN~\cite{yan2023spatio}, (ii) LLM-based in-context learning (ICL) methods like LLM-Mob~\cite{wang2023would}, PromptRec~\cite{wu2024could}, LightPROF~\cite{ao2025lightprof}, Refine-POI~\cite{li2025refine}, and (iii) supervised fine-tuned (SFT) LLMs like LLM4POI~\cite{li2024large}, GenUP~\cite{wongso2024genup}, GNPR-SID~\cite{10.1145/3711896.3736981}, GA-LLM~\cite{liu2025geography}. 
Each experiment is repeated with five random seeds, and we report the average performance. Details of the baselines are provided in Appendix A.5.

\subsection{Performance Analysis (RQ1)}
We compare our approach with a diverse set of baselines in Table~\ref{tab:overall_performance}. 
Graph-based models such as STHGCN rely primarily on structural transition patterns and thus show limited performance in next POI recommendation, where semantic understanding of user preferences is crucial.
Among LLM-based methods, ICL approaches with fixed prompts (e.g., LLM-Mob and PromptRec) achieve moderate gains by incorporating textual reasoning, but their static prompt designs limit adaptability. 
LightPROF further injects knowledge graph information into prompts and attains competitive performance without fine-tuning; however, its results remain consistently below optimization-based methods across all datasets, suggesting that fixed prompt structures constrain generalization to diverse urban contexts.
We further examine supervised fine-tuning (SFT) based methods. GenUP underperforms due to information loss from discarding long trajectories, while LLM4POI and Refine-POI improve performance through retrieval or reinforcement-based fine-tuning at the cost of additional training complexity.
GA-LLM achieves strong results via task-specific optimization and ranks as the runner-up in most settings.
In contrast, Prompt-as-Policy consistently achieves the best performance across all datasets and metrics, outperforming both static ICL methods and SFT-based approaches, without requiring any supervised fine-tuning.
These results demonstrate that dynamically learning how to select, organize, and compose evidences at the prompt level is more effective than static prompting or parameter-level optimization for next POI recommendation.

\subsection{Cold-start Analysis (RQ2)}

User activity level strongly affects recommendation performance: highly active users provide richer histories and more predictable mobility patterns, while inactive users represent challenging cold-start scenarios.
Following prior work~\cite{wongso2024genup,li2025refine}, we partition users into inactive, normal, and very active groups based on training trajectory counts. Specifically, the bottom 30\% of users ranked by trajectory count are defined as inactive, the top 30\% as very active, and the remaining users as normal. 
As shown in Table~\ref{tab:coldstart_performance}, inactive users have extremely limited historical data (e.g., 1.9 trajectories on average in NYC), confirming the severity of the cold-start setting.

Table~\ref{tab:coldstart_performance} further reports performance across user groups.
Static-prompt LLM methods achieve moderate gains but remain constrained under severe sparsity, while LightPROF benefits from KG-based evidence injection yet lacks adaptability due to fixed prompt structures. 
SFT-based methods, e.g., GNPR-SID, and GA-LLM, perform well for normal and very active users but are not explicitly optimized for cold-start scenarios and show limited robustness under extreme sparsity.
In contrast, our proposed Prompt-as-Policy consistently achieves the best performance on inactive users across all datasets, improving Acc@1 over the baseline by 12.02\%, 11.93\%, and 11.67\% on NYC, TKY, and CAL, respectively. 
Moreover, it maintains competitive performance for normal and very active users, indicating that the learned prompt policy adapts effectively to varying activity levels without any supervised fine-tuning.

\subsection{Sensitivity and Policy Analysis (RQ3)}

\begin{figure}[t]
    \centering
    \begin{subfigure}{0.48\linewidth}
        \centering
        \includegraphics[width=\linewidth]{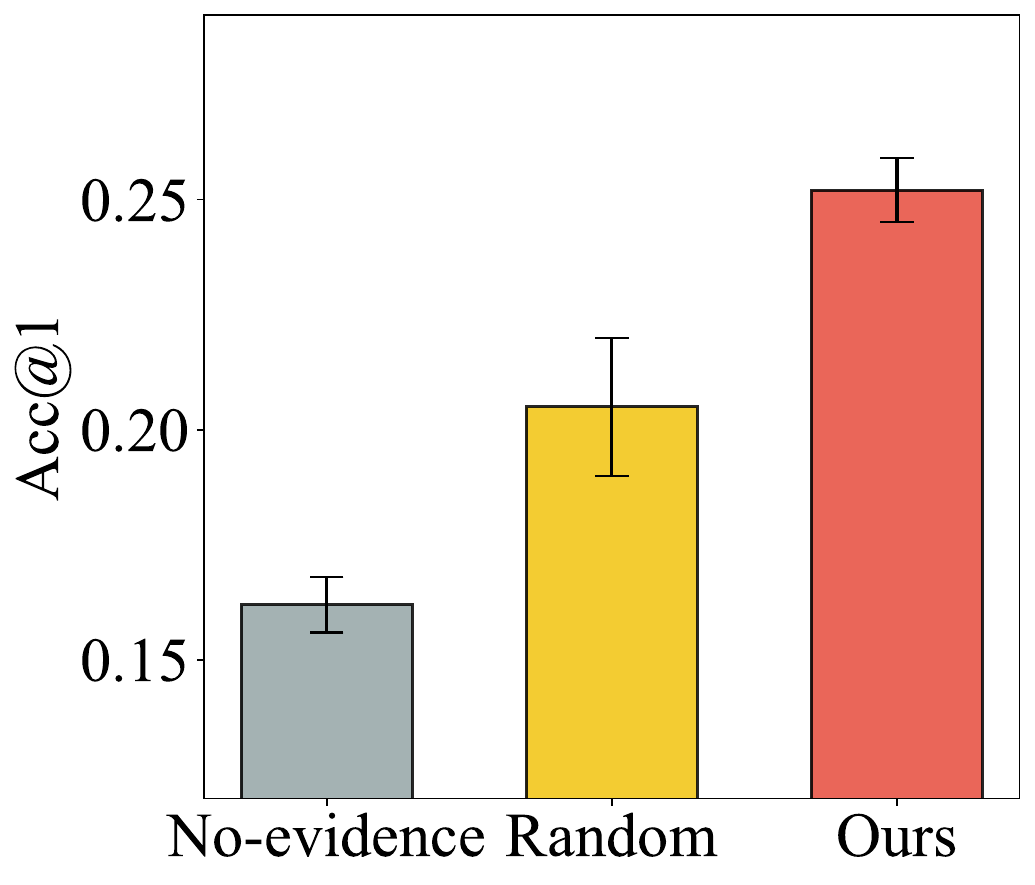}
        \caption{Impact of evidence selection}
        \label{fig:selection}
    \end{subfigure}
    \hfill
    \begin{subfigure}{0.48\linewidth}
        \centering
        \includegraphics[width=\linewidth]{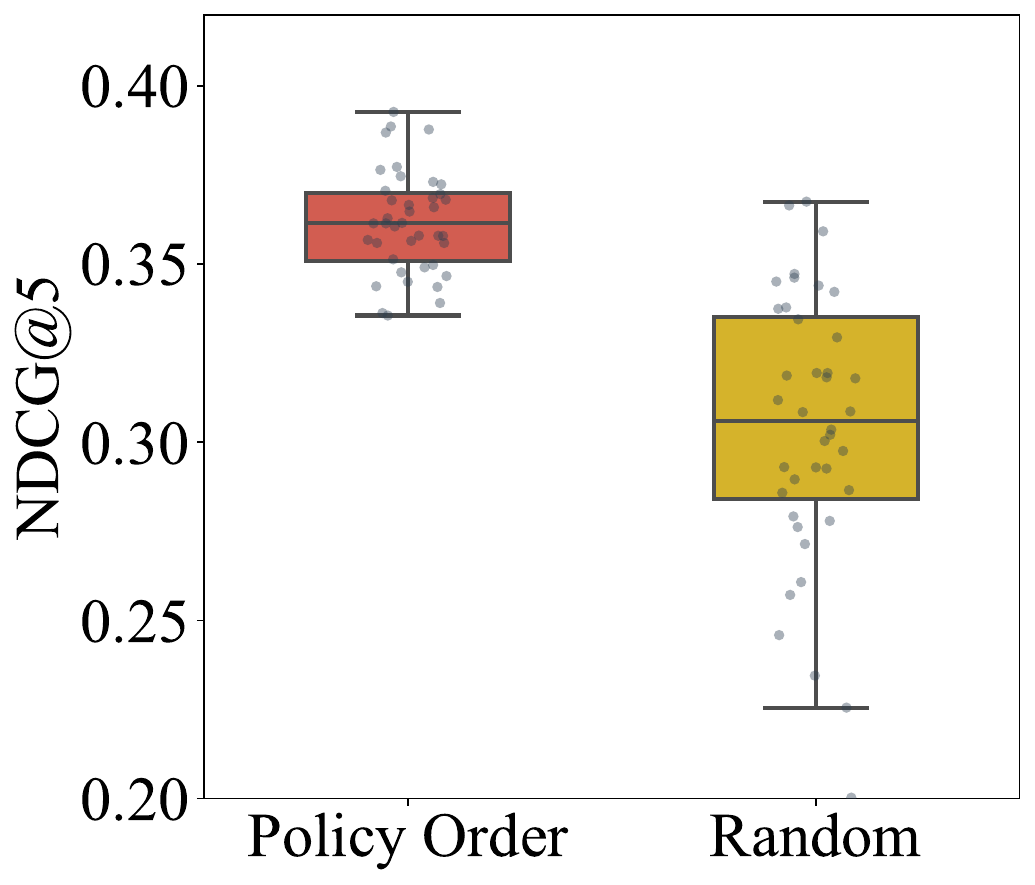}
        \caption{Impact of evidence ordering}
        \label{fig:ordering}
    \end{subfigure}

    \medskip 

    \begin{subfigure}{0.48\linewidth}
        \centering
        \includegraphics[width=\linewidth]{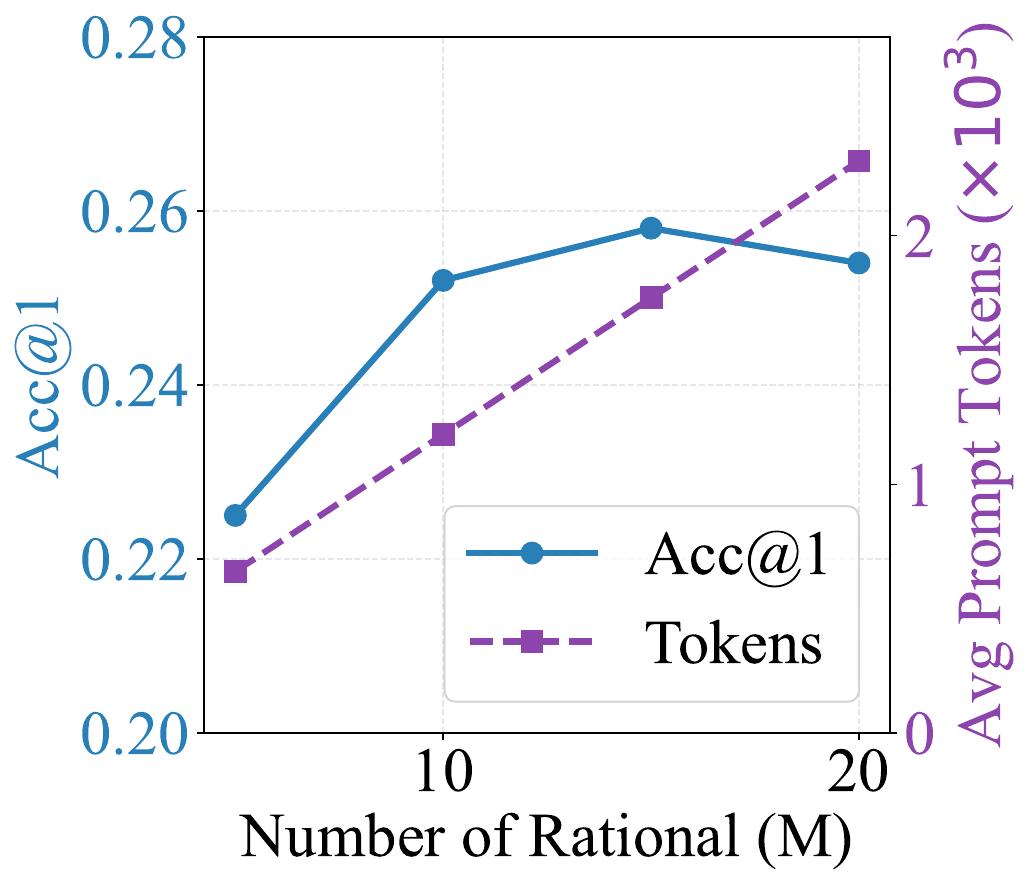}
        \caption{Impact of rational cap}
        \label{fig:cardinality}
    \end{subfigure}
    \hfill
    \begin{subfigure}{0.48\linewidth}
        \centering
        \includegraphics[width=\linewidth]{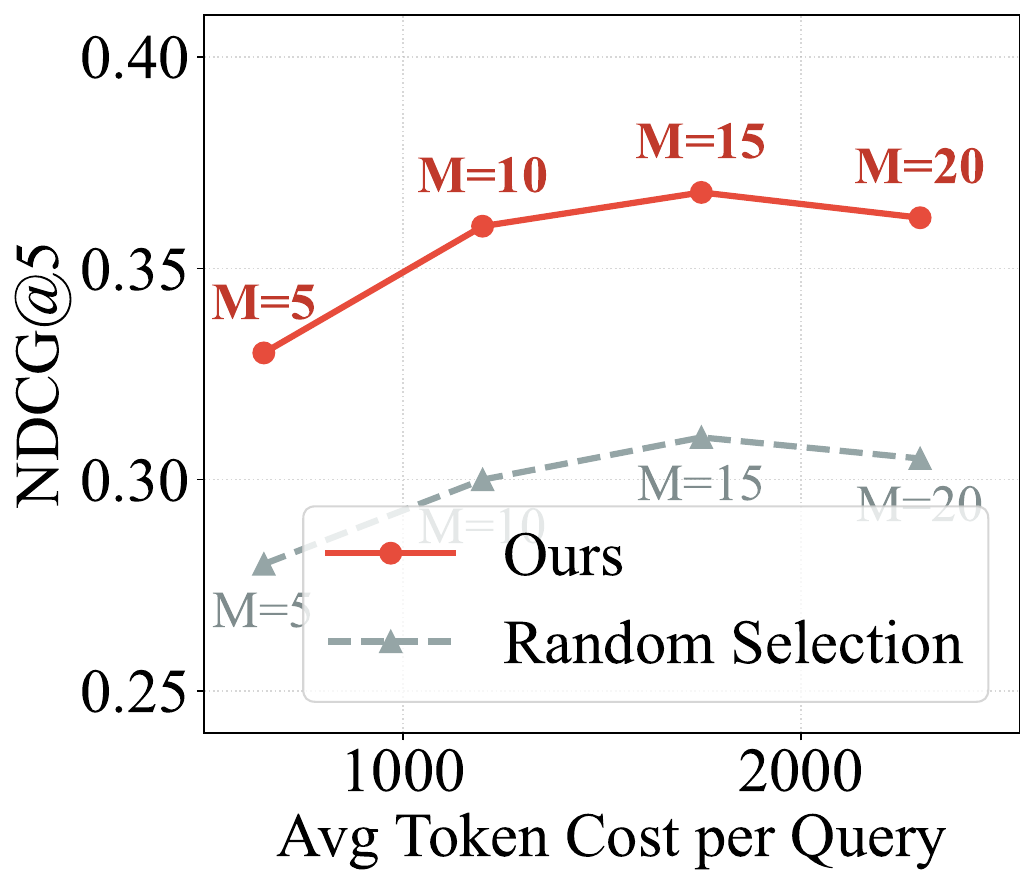}
        \caption{Average token cost}
        \label{fig:pareto}
    \end{subfigure}
    \caption{Sensitivity and policy analysis on the NYC dataset.} 
    \label{fig:rq3_analysis}
\end{figure}

To better understand the effectiveness of the learned prompt policy, we conduct sensitivity and policy analysis on evidence selection, evidence ordering, and rationale-cap control in prompt construction, and additionally report the associated token cost. Figure.~\ref{fig:rq3_analysis} shows the results in the NYC dataset for inactive users, the trends in TKY and CAL are similar. As shown in Figure.~\ref{fig:selection}, replacing the learned policy with no-evidence or random selection leads to a clear drop in Acc@1, indicating that adaptively selecting evidence from the same evidence pool is essential for effective prompting. Figure.~\ref{fig:ordering} further isolates the effect of ordering by keeping the evidence set fixed and only permuting their order, where the learned policy order consistently achieves higher NDCG@5 than random ordering. Figure.~\ref{fig:cardinality} varies the rationale cap and shows that performance improves and then saturates as M increases, while the average prompt tokens grow monotonically, revealing a practical accuracy–cost trade-off. Finally, Figure.~\ref{fig:pareto} summarizes this trade-off by plotting NDCG@5 against the average token cost per query, where the learned policy achieves better ranking quality under comparable token budgets.

\subsection{Ablation Study and Backbone Analysis (RQ4)}

\begin{figure}[t]
    \centering
    \begin{subfigure}{0.48\linewidth}
        \centering
        \includegraphics[width=\linewidth]{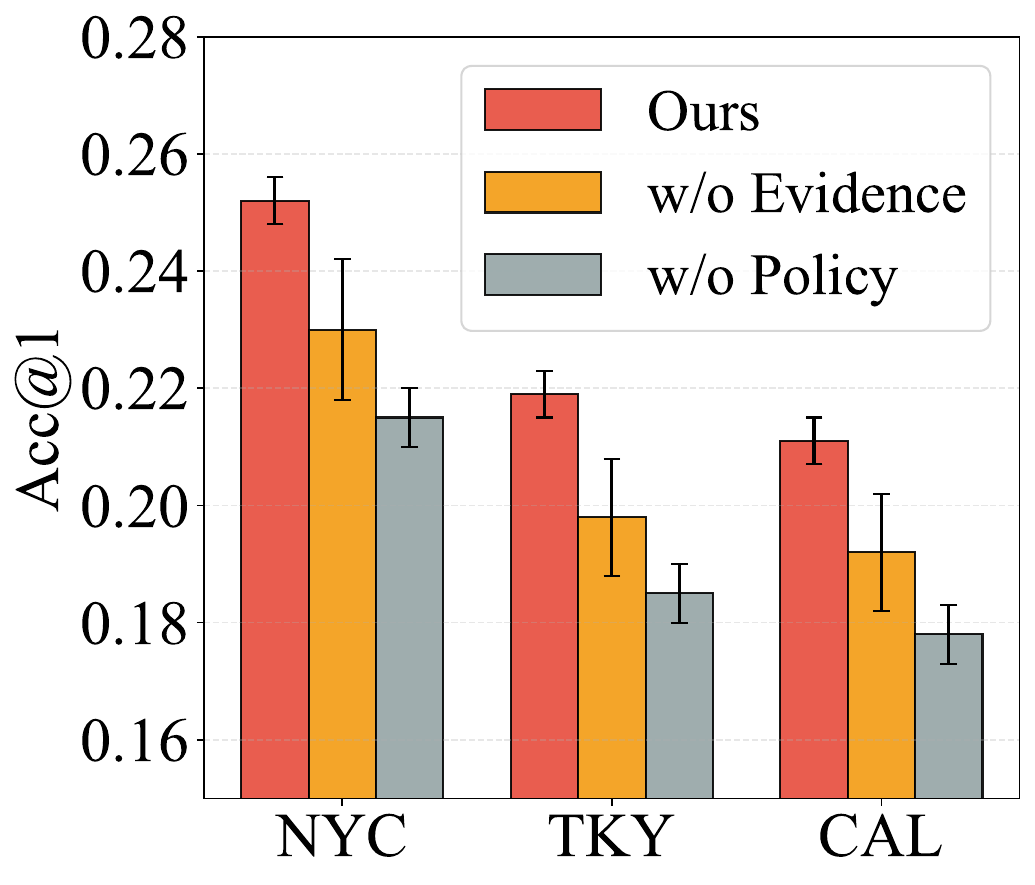}
        \caption{Ablation across datasets}
        \label{fig:ablation}
    \end{subfigure}
    \hfill
    \begin{subfigure}{0.48\linewidth}
        \centering
        \includegraphics[width=\linewidth]{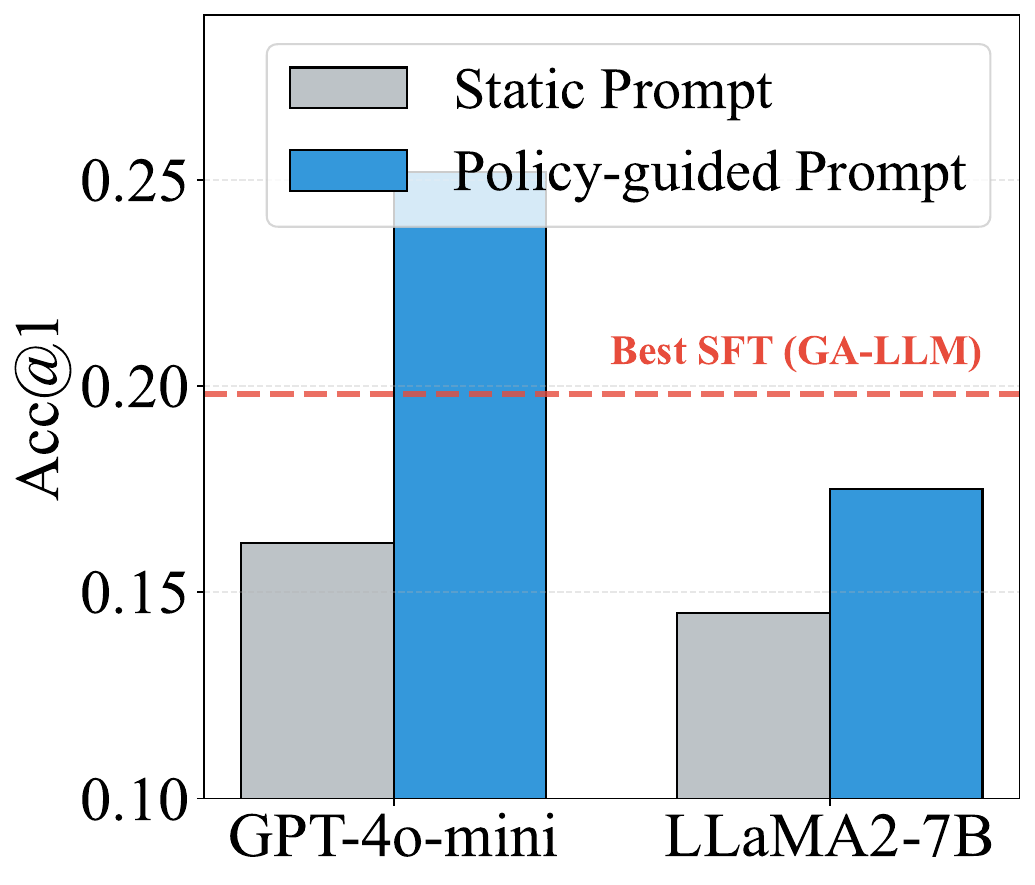}
        \caption{Backbone robustness}
        \label{fig:backbone}
    \end{subfigure}

    \caption{Ablation study and backbone analysis on the NYC dataset.}
    \label{fig:rq4_analysis}
\end{figure}

To better understand the sources of the gains in the proposed Prompt-as-Policy, we conduct component ablations and examine performance consistency across different backbone LLMs. 
As shown in Figure.~\ref{fig:ablation}, removing policy-guided prompt construction ($w/o\ Policy$), which discards the learned prompt policy and directly ranks candidates in $\mathcal{C}$ using heuristic scoring, causes the largest drop in Acc@1. 
This indicates that policy-guided prompt construction is particularly beneficial in cold-start settings, as it can flexibly extract the most relevant information from limited user histories, thereby improving recommendations without requiring SFT. 
In addition, removing evidence mining ($w/o\ Evidence$), where evidences are randomly selected for prompt construction instead of being mined via path sampling and candidate discovery, also degrades performance, suggesting that grounding the frozen LLM with informative KG evidences is essential. 
Figure.~\ref{fig:backbone} compares our policy-guided prompting against static prompting on both \texttt{gpt-4o-mini} and \texttt{LLaMA2-7B}, alongside the strongest SFT baseline (GA-LLM). Notably, \texttt{gpt-4o-mini} with static prompts underperforms the LLaMA-based SFT baseline. This indicates that a latest backbone alone is insufficient for SOTA performance in cold-start settings. However, applying our learned policy boosts \texttt{gpt-4o-mini} to 0.252, significantly surpassing all baselines. Furthermore, our policy consistently improves the \texttt{LLaMA2-7B} (+20.7\%), confirming that the performance gains primarily stem from the policy-guided prompt optimization rather than the backbone's capability.

\section{Conclusion}
In this work, we revisit the necessity of SFT for LLM-based next POI recommendation under cold-start conditions. We propose Prompt-as-Policy, a reinforcement-guided prompting framework that dynamically constructs evidence-based prompts over knowledge graphs. Unlike static prompting or fine-tuned models, our method keeps the LLM frozen as a reasoning engine and instead learns a contextual bandit policy that adaptively decides which evidences to include, how many to retain, and how to organize them within prompts. Extensive experiments on three real-world Foursquare datasets show that Prompt-as-Policy consistently outperforms both SFT-based and static-prompt baselines, particularly in cold-start scenarios. Moreover, comparable results between LLMs variants confirm that the performance gain primarily stems from the learned prompt policy rather than differences in the LLMs. For future work, we plan to extend the Prompt-as-Policy paradigm beyond next POI recommendation to broader reasoning driven tasks, such as conversational recommendation, where policy-guided prompt optimization can further improve reasoning stability.

\bibliographystyle{named}
\bibliography{ijcai26}

\end{document}